\documentclass[%
reprint,
 amsmath,amssymb,
 prl,
]{revtex4-2}

\usepackage{graphicx}
\usepackage{dcolumn}
\usepackage{bm}
\usepackage{flushend}
\usepackage{hyperref}
\usepackage{color}
\hypersetup{colorlinks=true, citecolor=blue, urlcolor=blue, linkcolor=blue}
\begin{document}
\title{Modal dynamics in multimode optical fibers: An attractor of high-order modes}

\author{Weitao He} 
\author{Ruihuan Wu} 
\author{Weiyi Hong}\email{hongwy@m.scnu.edu.cn}
\author{Aiping Luo}\email{luoaiping@scnu.edu.cn}

\affiliation{Guangdong Provincial Key Laboratory of Nanophotonic Functional Materials and Devices, South China Normal University, Guangzhou 510631, P. R. China.
}

\date{\today}%
\begin{abstract}

Multimode fibers (MMFs) support abundant spatial modes and involve rich spatiotemporal dynamics, yielding many promising applications. Here, we investigate the influences of the number and initial energy of high-order modes (HOMs) on the energy flow from the intermediate modes (IMs) to the fundamental mode (FM) and HOMs. It is quite surprising that random distribution of high-order modes evolves to a stationary one, indicating the asymptotic behavior of orbits in the same attraction domain. By employing the Lyapunov exponent, we prove that the threshold of the HOMs-attractor is consistent with the transition point of the energy flow which indiactes the HOMs-attracotr acts as a "valve" in the modal energy flow. Our results provide a new perspective to explore the nonlinear phenomena in MMFs, such as Kerr self-cleaning, and may pave the way to some potential applications, such as secure communications in MMFs.
\end{abstract}
\maketitle


In recent years, multimode fibers (MMFs) are extensively investigated due to the great potential in spatial division multiplexing systems and multimode fiber lasers \cite{1,2}. Meanwhile, MMFs offer larger mode areas, support more spatial modes, and provide a new degree of freedom to control the optical field \cite{3}. MMFs involve complex spatiotemporal dynamics and intrinsic disorder, and they have been used to investigate a series of new nonlinear spatiotemporal dynamics phenomena, such as spatiotemporal mode-locking \cite{4,5,6,7},multimode solitons \cite{1,8,9}, intermodal four-wave mixing  \cite{10,11,12}, geometric parametric instability  \cite{13,14,15,16,17},  spatiotemporal modulation instability \cite{18}, and Kerr beam self-cleaning \cite{3,19,20,21,22,23}. These phenomena are of great interest not only from the point of view of basic science, but also in various practical applications \cite{24}.

As is known to all, optical field propagation in MMFs involves rich spatiotemporal dynamics and complex intermodal interactions \cite{2,25,26}. More theoretical studies and ways are needed for a more profound understanding of complex spatiotemporal dynamic behaviors \cite{25}. Control of physical phenomena in MMFs and their applications are in their infancy, opening opportunities to take advantage of complex nonlinear modal dynamics \cite{27}. So far, although the physical mechanism of the Kerr beam self-cleaning phenomenon is an open problem, researchers have introduced concepts of the hydrodynamic 2D turbulence \cite{21,28}, the thermalization and condensation in thermodynamics \cite{29} to provide a very constructive explanation. These research methods have offered some interesting ways and ideas for exploring the dynamic process in MMFs and opening up a new visual field.

The studies of E. V. Podivilov et al. showed that in the process of the mode energy transmission, the energy of the intermediate modes (IMs) flows into the fundamental mode (FM) and high-order modes (HOMs), which is analogous to the hydrodynamic 2D turbulence \cite{21} and therefore contains rich physical phenomena. The HOMs with much lower energy compared to those of FM and IMs acts as the background noise but plays an important role in the modal energy flow. In this letter, we systematically study the energy flow in the graded-index MMFs, and find that small mode number and tiny energy of the HOMs trigger the energy flow from the IMs to the FM and the HOMs. Quite surprisingly, the energy flow exhibits an attractor of the HOMs: random mode-energy distribution of HOMs evolves to the stationary one, and such stationary distribution of the HOMs depends on the initial energy distribution of the FM and IMs. We further adopt the Lyapunov exponent to demonstrate the forming process of the HOMs-attractor, and find that initial energy proportion of HOMs determines the threshold of the HOMs-attractor. Interestingly, this threshold is consistent with the transition point of the energy outflow from the IMs, suggesting that the HOMs-attractor is a "valve" for the energy flow.


In MMFs, the description of pulse propagation often involves spatial and temporal dynamics at the same time. The modal decomposition translates to higher accuracy, and makes a clear interpretation of spatiotemporal dynamics in MMFs. We numerically solve the generalized nonlinear Schrödinger equation (MM-GNLSE) \cite{2,30,31} to explore the process of modal dynamics in the following. The evolution of the electric field temporal envelope of the spatial mode p versus the propagation distance z can be written as:
\begin{figure}
    \centering
    \includegraphics[scale = 0.42]{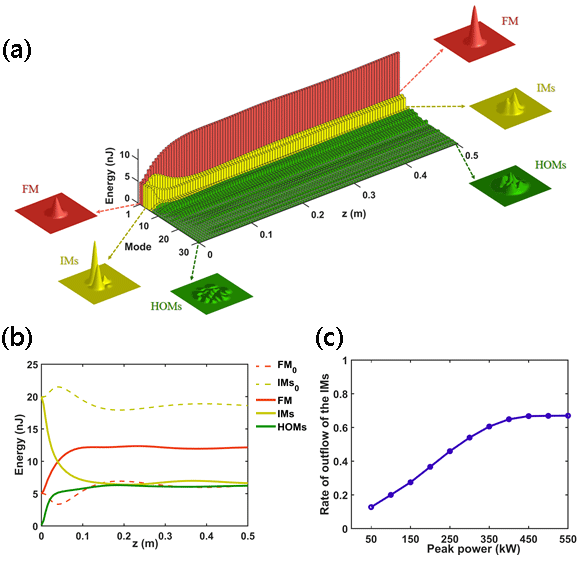}
    \caption{(a) Evolutions of the fundamental mode (FM), intermediate modes (IMs) and higher-order modes (HOMs) upon the propagation distance z. Illustrations represent the beam intensity patterns at the input and output end of the fiber, respectively. (b) Energy flows of the FM, IMs and HOMs where z is 0.5 m. Dotted curves indicate the case that there is no HOMs in the initial field. (c) The rate of outflow of the IMs at the output end of the fiber versus the peak power.}
    \label{figure_1}
\end{figure}

\begin{equation}
\begin{aligned}
& \partial _{z}A_{p}(z,t) = \\
& i\delta \beta_{0}^{(p)}A_{p} -\delta \beta _{1}^{(p)}\partial_{t}A_{P}+  \sum_{m = 2}^{N_{d}}i^{m+1}\frac{\beta_{m}^{(p)}}{m!}\partial_{t}^{m}A_{p} +\\ 
& i\frac{n_{2}\omega_{0}}{c}(1+\frac{i}{\omega_{0}}\partial_{t})\sum^{N}_{l,m,n}[(1-f_{R})S_{plmn}^{K}A_{l}A_{m}A_{n}^{\ast} +\\ 
& f_{R}S_{plmn}^{R}A_{l}\int_{-\infty}^{t}d\tau h_{R}(\tau)A_{m}(z,t-\tau)A_{n}^{\ast}(z,t-\tau)],
\end{aligned}
\end{equation}

\noindent where the first term, the second term and the third term on the right hand side indicate the propagation constant mismatch, the modal dispersion and the higher order dispersion effects, respectively. In the fourth term, $n_{2}$ represents the nonlinear refractive index, $f_{R}$ shows the fractional contribution of the Raman effect ( $f_{R}$ = 0.18 for silica glass fibres), $h_{R}$ is the Raman response of the fiber medium, $S_{plmn}^{K}$ and $S_{plmn}^{R}$ are the mode overlap factors responsible for the Raman and Kerr effect severally \cite{31}.

\par We launched Gaussian pulses with with the duration of 50 fs, the wavelength of 1064 nm, and the initial total energy of 25 nJ into a 0.5-m-long graded-index MMF. The core radius of the fiber is 25 $\mu m$, and the value of the refractive index difference is 0.013. In the dynamics of the modal energy flow, the energy of the IMs flows into the FM and the HOMs as shown in Fig.~\ref{figure_1}. In the following discussion, we regard the second to the fifth modes as the IMs and the next high-order modes as the HOMs. Firstly, we set 
the equal energy to the first five modes and the initial average energy ratio of $E_{AEH}/E_{AEF}$ to 1/1000 (marked the $E_{AEH}$ and $E_{AEF}$ as the initial average energy of the HOMs and the initial average energy of the first five modes, severally). Figure \ref{figure_1}(a) shows the energy exchanges among the 30 spatial modes with the increase of propagation distance. These three-dimensional beam intensity models represent the FM, IMs and HOMs at the input and output end of the fiber, respectively. In Fig.~\ref{figure_1}(b), the solid curves and dotted curves represent the energy flows where the HOM number is 25 and 0 in the initial field, severally. Obviously, abundant energy flows from the IMs to the FM and HOMs when the HOMs are excited in the initial field. Figure \ref{figure_1}(c) presents the relationship between the peak power and the rate of outflow of the IMs (the energy flowing out of the IMs divided by the initial energy of the IMs). The rate rises up and tends to saturate at the value of the peak power is around 450 kW.

\begin{figure}[b]
    \centering
    \includegraphics[scale = 0.27]{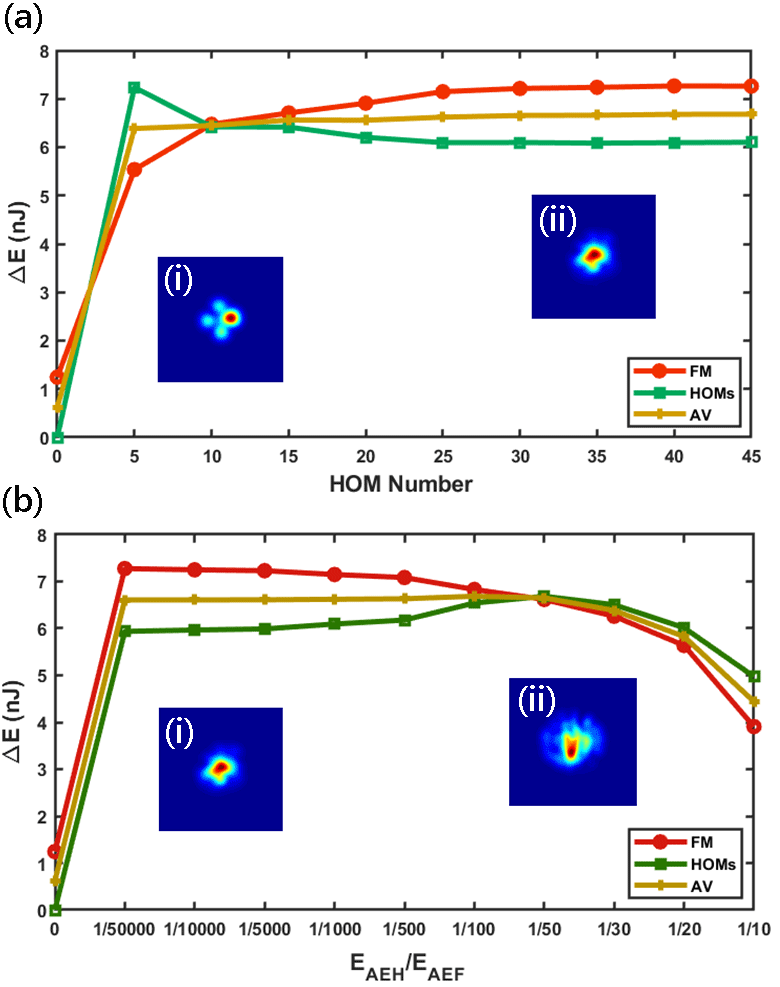}
    \caption{(a) Energy difference between the output and input energies versus the HOM number (the ratio of the $E_{AEH}$ to the $E_{AEF}$ is 1/1000). Inset: (i) and (ii) correspond to the output beam profiles where the HOM number is 0 and 45, respectively. (b) Energy difference between the output and input energies for the varying ratios of the $E_{AEH}$ to the $E_{AEF}$ (the HOM number is set to 30). Inset: (i) and (ii) correspond to the output beam profiles where the ratio is 1/50,000 and 1/10, severally. AV is the averge value of the energy varitons of the FM and HOMs. In order to keep the peak power and the initial energy of the first five modes unchanged, we increased the total input energy and pulse width appropriately.}
    \label{figure_2}
\end{figure}

\begin{figure*}[t]
    \centering
    \includegraphics[scale = 0.28]{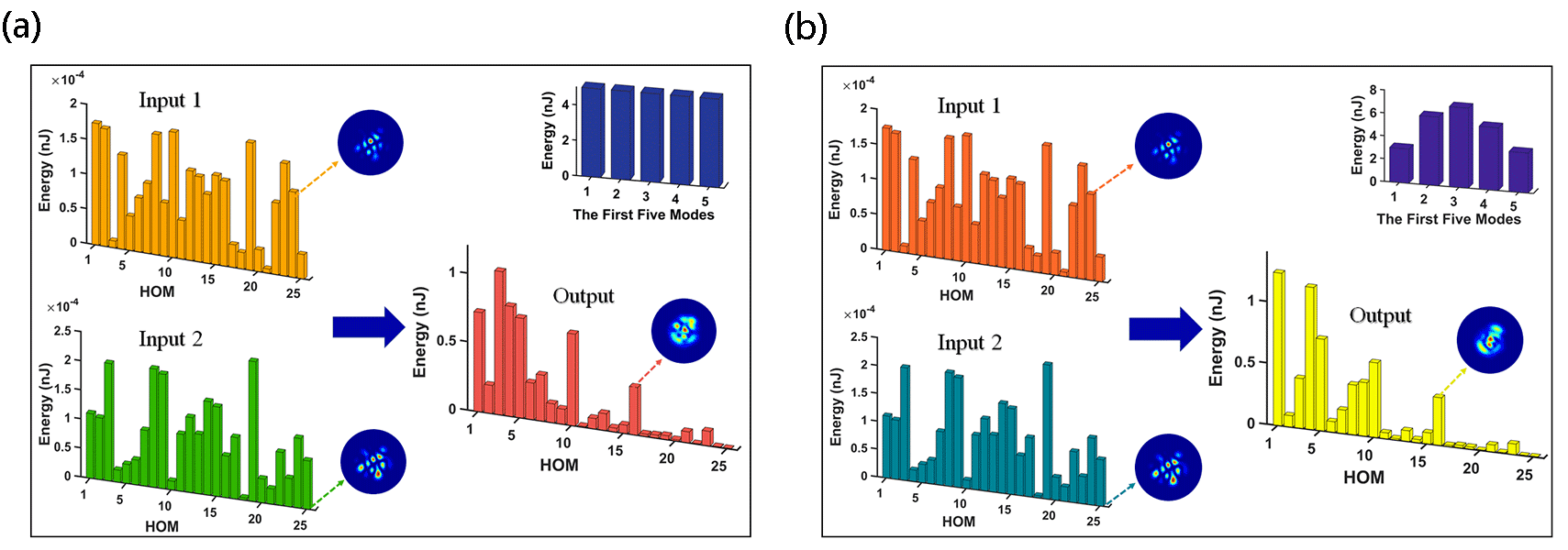}
    \caption{The numerical evolution results of the 25 HOMs with different initial energy distributions of the first five modes (peak power is also 500 kW). (a) Input 1 and input 2 represent two different random initial energy distributions of the HOMs. The outputs of input 1 and input 2 are the same and the three beam profiles in this figure correspond to the input 1, input 2 and output, respectively. (b) Same conditions as (a) except for the initial energy distribution of the first five modes (0.12, 0.24, 0.28, 0.22, 0.14).}
    \label{figure_3}
\end{figure*}

\par  Since the noise background formed by the HOMs leads to the energy flow from IMs to the FM and the HOMs, the mode number and the energy proportion of the HOMs would naturally influence this process. We show in Fig. \ref{figure_2}(a) the difference between the output and input energies for both FM and HOMs as functions of the HOM number. The average of them implies the total energy flow from the IMs. It is interesting that very small number of HOMs significantly triggers the energy flow, and the energy flow from the IMs saturates immediately as the HOMs number increases. The energy flow to the HOMs exhibits a slow decrease after its sharp increase. The insets (i) and (ii) representing the spot of the total output field depict the beam cleaning improved by the HOMs. We keep the HOM number of $30$ which is within the saturation region as shown in Fig. \ref{figure_2}(a), and investigate the influence of the initial energy of the HOMs on the energy flow. Figure \ref{figure_2}(b) show that extremely weak noise background of the HOMs is enough to trigger the energy flow. As the initial energy of the HOMs increases, the energy flow from the IMs keeps unchangeable and significantly declines when initial energy propotion of the HOMs reaches a certain value of $E_{AEH}/E_{AEF}=1/50$. It is worth mentioning that the energy flow to the HOMs experiences a rise and reach its maximum also at $E_{AEH}/E_{AEF}=1/50$. We will see below that it is the transformation point of the HOMs. The insets (i) and (ii) also implies that weaker energy of the HOMs facilitates the beam cleaning.

\par Previous studies for the modal energy flow always focused on the energy flow to the FM, i.e., the beam cleaning, due to the its applications of bright FM beam generation. Quite different from the previous studies, we focus on the evolution of the energy distribution of the HOMs in the modal energy flow, and find a very surprising phenomenon: \textit{the dynamics of the HOMs exhibit an attractor}. Fig. \ref{figure_3}(a) illustrates that two \textit{random} initial energy distributions of the HOMs (respectively marked as Input 1 and Input 2) evolve to the same distribution after the propagation of the 0.5-m long fiber. In this simulation, the energies of the FM and IMs (the lowest five modes) are assumed to be the same value of 5 nJ and the energy proportion of the HOMs is set to $E_{AEH}/E_{AEF}=1/50000$. To avoid coincidence, we have examined this process by multiply randomizing the distribution of HOMs. In other words, random distribution of the HOMs will evolve to a stationary one. Note that the initial energies of the FM and IMs and initial average energy of the HOMs are fixed in the simulation, therefore we could reasonably conjecture the existence of an attractor of HOMs in the phase space constructed by the mode energies. We further perform the simulation with the same parameters as those in Fig. \ref{figure_3}(a) but for different energy distribution of the lowest five modes . As expected, random distribution of the HOMs in Fig. \ref{figure_3}(b) evolves to the attractor different from that in Fig. \ref{figure_3}(a).Therefore, it can be concluded that such attractor of HOMs depends on the initial energy distribution of the FM and IMs, and also the total energy proportion the HOMs.

\begin{figure}[b]
    \centering
    \includegraphics[scale = 0.175]{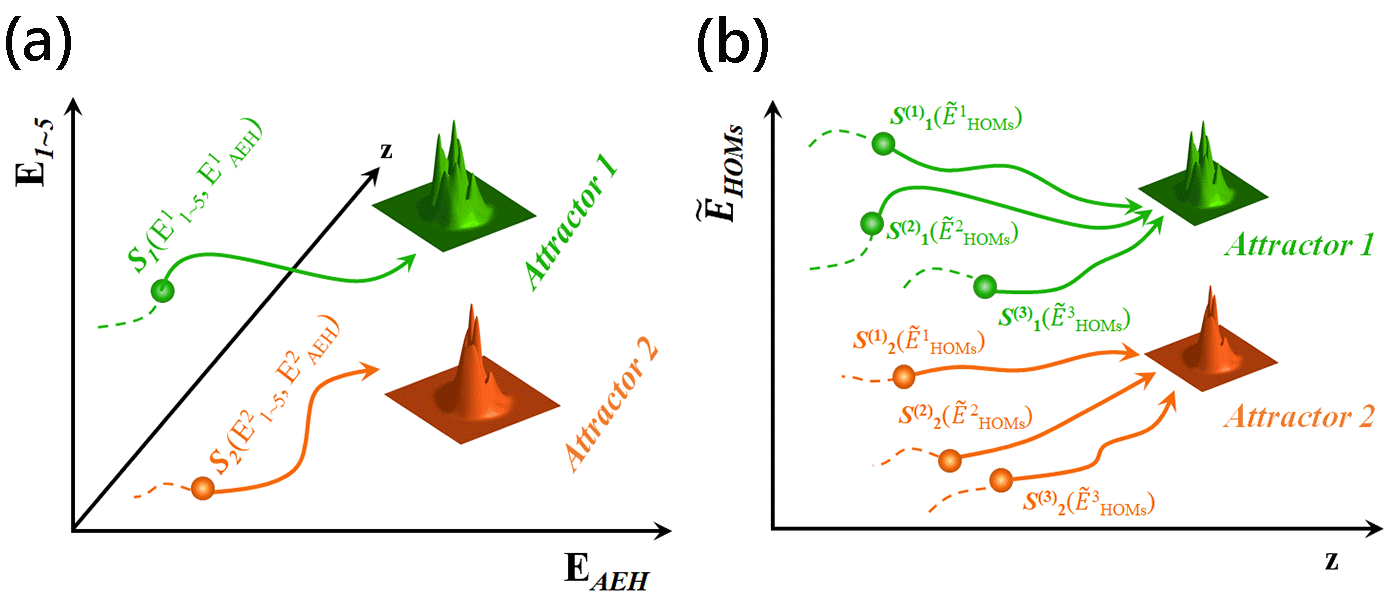}
    \caption{Schematic diagram of the attractors of HOMs. (a) A six-dimensional phase space consisting of the energies of the first five modes ($E_1$, $E_2$, $E_3$, $E_4$, $E_5$) and the HOMs ($E_{AEH}$). Two different states ($S_1$ and $s_{2}$) respectively evolve along their own orbits towards corresponding attractors which are determined by the initial energy distributions of the first five modes. (b) Three different sub-states ($S^{(1)}_1$, $S^{(2)}_1$, $S^{(3)}_1$) which have the same the initial energy distributions of the first five modes evolve towards the same attractor (Attractor 1) and the orther three evolve towards another attractor (Attractor 2).}
    \label{figure_4}
\end{figure}

\begin{figure*}[t]
    \centering
    \includegraphics[scale = 0.28]{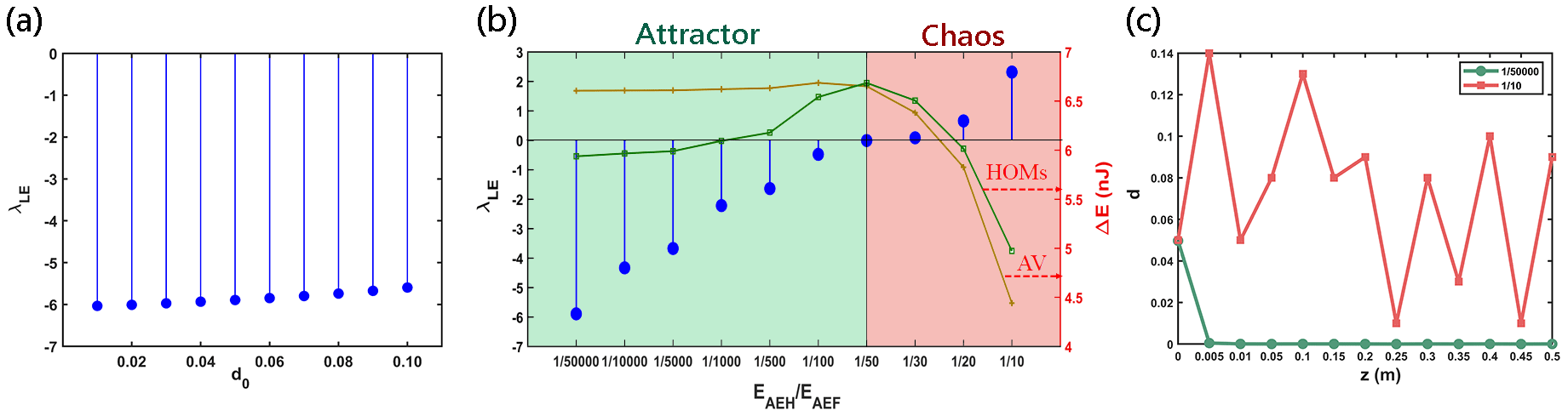}
    \caption{(a) Correlation between the Lyapunov exponent and the different initial distance of two orbits. (b) The Lyapunov exponent and the energies difference of the HOMs and AV (same as Fig.~\ref{figure_3}(b)) for the varying ratios of the $E_{AEH}$ to the $E_{AEF}$). This area is divided to the Attractor and Chaos based on the positive and negative values of the Lyapunov exponent. (c) The normalized distance d of the two orbits upon the propagation distance z for the attractor ($1/50000$) and chaos (1/10).}
    \label{figure_5}
\end{figure*}

\par Based on the above numerical results for the evolution of the HOMs which evolve the behavior of the attractors, we would like to further schematically demonstrate the concept of such attractor in the phase space constructed by the mode energies. The attractor is regarded as the asymptotic behavior of orbits \cite{32}, which originates from the topological properties of differential equations \cite{32}.  Figure \ref{figure_4}(a) depicts the orbits for two attractors in the phase space $\pounds$[$E_1$, $E_2$, $E_3$, $E_4$, $E_5$, $E_{AEH}$]. $S_1(E^1_{1-5},E^1_{AEH})$ and $S_2(E^2_{1-5},E^2_{AEH})$ are two different states in $\pounds$. They evolve along their own inherent orbits (marked by the arrows) and tend asymptotically to the corresponding attractors. Although these two states in the actual light propagation in the fiber may be the initial conditions, they could be the intermediate states in the orbits, rather than being regarded as the beginning of the orbits. It should be mentioned that any state in $\pounds$ can only evolve along an exclusive orbit to its corresponding attractor, and any two orbits never intersect in $\pounds$.

It could be vivid to see the behavior of the attractor in sub-space $l$[$\widetilde{E}_{HOMs}$] constructed by the mode energies of HOMs ($\widetilde{E}_{HOMs}$ represents the series of the energies of HOMs normalized by $E_{AEH}$), which is shown in Fig.~\ref{figure_4}(b). In $l$, three different sub-states $S^{(1)}_1(\widetilde{E}^1_{HOMs})$, $S^{(2)}_1(\widetilde{E}^2_{HOMs})$ and $S^{(3)}_1(\widetilde{E}^3_{HOMs})$, which are all belong to the state $S_1$ in $\pounds$ since their initial energies of the first five modes and the intial average energy of the HOMs are the same, will evolve asymptotically along their own sub-orbits to the same state (marked as Attractor1). And the same goes for another three sub-states $S^{(1)}_2(\widetilde{E}^1_{HOMs})$, $S^{(2)}_2(\widetilde{E}^2_{HOMs})$ and $S^{(3)}_2(\widetilde{E}^3_{HOMs})$.


\par Lyapunov exponent would be helpful to deeply investigate the behavior of attractor demonstrated above, which can be expressed as follows: 
\begin{equation}
        \lambda_{LE} =\lim_{L\rightarrow \infty} \frac{1}{L}\ln\frac{\left \| S^{(2)}_{z=L}-S^{(1)}_{z=L}\right \|_2}{\left \|S^{(2)}_{z=0}-S^{(1)}_{z=0} \right \|_2},
\end{equation}
where $S^{(1)}$ and $S^{(2)}$ are two sub-states in $\l$, and are both belong to the state $S$ in $\pounds$, and $L$ is the propagation distance of the fiber. $\|*\|_2$ denotes the $2$-norm in $\l$. When $\lambda_{LE}<0$, the system exhibits a behavior of attractor, otherwise the system is chaotic. Although $L$ should tend to infinite in the definition of $\lambda_{LE}$, it can be sufficiently long where the energy flow tend to be saturated in the practical simulations. We assume the parameters for the state $S$ as Fig. \ref{figure_3}(a) and calculate the spectrum of its Lyapunov exponent $\lambda_{LE}(d)$ with $d=\|S^{(2)}-S^{(1)}\|_2$ the $2$-norm, i.e., the distance of two initial sub-states. As shown in Fig.~\ref{figure_5}(a), $\lambda_{LE}$ is always near $-5.85$ with a slight rise as $d_{0}$ (initial distance of two adjacent sub-orbits) increases from $0.01$ to $0.1$. 

In Fig.~\ref{figure_5}(b), we further shift the state $S$ along the axis $E_{AEH}$ in the space of $\pounds$ shown in Fig.~\ref{figure_4}(a), and find that $\lambda_{LE}$ tends to zero and subsequently becomes positive as the increases of $E_{AEH}$. The diminishing of $|\lambda_{LE}|$ for the attractor means that the system need longer propagation distance to converge to the attractor. The result also shows that the system exist a transformation point of the attractor near $E_{AEH}/E_{AEF}=1/50$. The spatial beam self-cleaning in MMFs originates from a universal unstable attractor and once the critical state of the attractor is reached, the initial field will self-organizes into a stable state \cite{18,26}.
When the initial energy of the HOMs is very low, it can be regarded as a perturbation of the total initial field. Therefore, as the total initial field self-organizes into a stable state, the HOMs with random initial energy distribution will reach the same stable state (the orbits in the same attraction domain evolve into the same attractor). Conversely, when the initial energy of the HOMs is high, the field tends to a chaotic state due to the large disturbance. We also plot the energy flow from the IMs, and find that it keeps unchangeable within region where the attractor, and declines from the transformation point of the attractor, which implies that the HOMs-attractor is a "valve" for the energy flow from the IMs. We further calculate the evolution of the distance of two sub-states in the regions of attractor and chaos, respectively. As expected, the distance for the case of attractor reduces rapidly to zero as shown in Fig.~\ref{figure_5}(c), implying the fast merging of two adjacent sub-orbits, while that for the case of chaos fluctuates during propagation, implying the unpredictable behavior of the sub-orbits. 

In conclusion, we systematically investigate the mode dynamics in the graded-index MMFs, and focus on the energy flow from IMs to FM and HOMs. It is found that small mode number and tiny energy proportion of HOMs significantly trigger the energy flow, and the mode number exhibits seldom influence on the energy flow when it is higher than 30. It is found a surprising phenomenon that the evolution of the HOMs exists the behavior of attractor when the energy proportion of the HOMs is under a certain value: random distribution of the HOMs evolves to a stationary one, and the attractor depends on the initial distributions of the FM and the IMs. Our results show that the HOMs-attractor acts as a "valve" in the modal energy flow, which may provide a new perspective to explore the nonlinear phenomena in MMFs, such as Kerr self-cleaning. We further adopt Lyapunov exponent to quantitatively discuss the behavior of the attractor of HOMs. Base on such consistent one-to-one match between the input of FM and IMs and the output of the attractor of HOMs, this phenomenon might be applied on the area of secure communications.

\section{Acknowledgements}
This work was supported by the National Natural Science Foundation of China (Grant NO.  92050101, 61875058, 11874019) and the Natural Science Foundation of Guangdong Province under Grant No. 2019A1515011172.


\begin{thebibliography}{0}%
\makeatletter
\providecommand \@ifxundefined [1]{%
 \@ifx{#1\undefined}
}%
\providecommand \@ifnum [1]{%
 \ifnum #1\expandafter \@firstoftwo
 \else \expandafter \@secondoftwo
 \fi
}%
\providecommand \@ifx [1]{%
 \ifx #1\expandafter \@firstoftwo
 \else \expandafter \@secondoftwo
 \fi
}%
\providecommand \natexlab [1]{#1}%
\providecommand \enquote  [1]{``#1''}%
\providecommand \bibnamefont  [1]{#1}%
\providecommand \bibfnamefont [1]{#1}%
\providecommand \citenamefont [1]{#1}%
\providecommand \href@noop [0]{\@secondoftwo}%
\providecommand \href [0]{\begingroup \@sanitize@url \@href}%
\providecommand \@href[1]{\@@startlink{#1}\@@href}%
\providecommand \@@href[1]{\endgroup#1\@@endlink}%
\providecommand \@sanitize@url [0]{\catcode `\\12\catcode `\$12\catcode
  `\&12\catcode `\#12\catcode `\^12\catcode `\_12\catcode `\%12\relax}%
\providecommand \@@startlink[1]{}%
\providecommand \@@endlink[0]{}%
\providecommand \url  [0]{\begingroup\@sanitize@url \@url }%
\providecommand \@url [1]{\endgroup\@href {#1}{\urlprefix }}%
\providecommand \urlprefix  [0]{URL }%
\providecommand \Eprint [0]{\href }%
\providecommand \doibase [0]{https://doi.org/}%
\providecommand \selectlanguage [0]{\@gobble}%
\providecommand \bibinfo  [0]{\@secondoftwo}%
\providecommand \bibfield  [0]{\@secondoftwo}%
\providecommand \translation [1]{[#1]}%
\providecommand \BibitemOpen [0]{}%
\providecommand \bibitemStop [0]{}%
\providecommand \bibitemNoStop [0]{.\EOS\space}%
\providecommand \EOS [0]{\spacefactor3000\relax}%
\providecommand \BibitemShut  [1]{\csname bibitem#1\endcsname}%
\let\auto@bib@innerbib\@empty
\end{thebibliography}%


\begin{thebibliography}{99}

\bibitem{1} L. G. Wright, W. H. Renninger, D. N. Christodoulides, and F. W. Wise, Opt. Express.{\bf 23}, 3492 (2015).

\bibitem{2} P. Horak and F. Poletti, Recent Prog. Opt. Fiber Res. ch. 1, pp. 3–24.(2012).

\bibitem{3}Z. Liu, L. G. Wright, D. N. Christodoulides, and F. W. Wise, Opt. Lett.{\bf 41}, 3675 (2016).

\bibitem{4}	H. Qin, X. Xiao, P. Wang, and C. Yang, Opt. Lett. {\bf 43}, 1982 (2018).

\bibitem{5}	Y. Ding, X. Xiao, P. Wang, and C. Yang, Opt. Express {\bf 27}, 11435 (2019).

\bibitem{6}	L. G. Wright, P. Sidorenko, H. Pourbeyram, Z. M. Ziegler, A. Isichenko, B. A. Malomed, C. R. Menyuk, D. N. Christodoulides, and F. W. Wise, Nat. Phys. {\bf 16}, 565 (2020).

\bibitem{7}	Y. Ding, X. Xiao, K. Liu, S. Fan, X. Zhang, and C. Yang, Phys. Rev. Lett. {\bf 126}, 93901 (2021).

\bibitem{8}	W. H. Renninger and F. W. Wise, Nat. Commun. {\bf 4}, 1719, (2013).

\bibitem{9}	L. Rishøj, B. Tai, P. Kristensen, and S. Ramachandran, Optica {\bf6}, 304 (2019).

\bibitem{10}	E. Nazemosadat, H. Pourbeyram, and A. Mafi, J. Opt. Soc. Am. B {\bf33}, 144 (2016).

\bibitem{11}	R. Dupiol, A. Bendahmane, K. Krupa, A. Tonello, M. Fabert, B. Kibler, T. Sylvestre, A. Barthelemy, V. Couderc, S. Wabnitz, and G. Millot, Opt. Lett. {\bf42}, 1293 (2017).

\bibitem{12}	A. Bendahmane, K. Krupa, A. Tonello, D. Modotto, T. Sylvestre, V. Couderc, S. Wabnitz, and G. Millot, J. Opt. Soc. Am. B {\bf35}, 295 (2018).

\bibitem{13}	K. Krupa, A. Tonello, A. Barthélémy, V. Couderc, B. M. Shalaby, A. Bendahmane, G. Millot, and S. Wabnitz, Phys. Rev. Lett. {\bf116}, 183901 (2016).

\bibitem{14}	C. Mas Arabí, A. Kudlinski, A. Mussot, and M. Conforti, Phys. Rev. A {\bf97}, 1 (2018).

\bibitem{15}	H. E. Lopez-Aviles, F. O. Wu, Z. Sanjabi Eznaveh, M. A. Eftekhar, F. Wise, R. Amezcua Correa, and D. N. Christodoulides, APL Photonics {\bf4}, 022803 (2019).

\bibitem{16}	M. Conforti, C. Mas Arabi, A. Mussot, and A. Kudlinski, Opt. Lett. {\bf42}, 4004 (2017).

\bibitem{17}	W. HE, J. DAI, Q. MA, A. LUO, and W. HONG*, Opt. Express.{\bf 29}, 11353 (2021).

\bibitem{18}	L. G. Wright, Z. Liu, D. A. Nolan, M.-J. Li, D. N. Christodoulides, and F. W. Wise, Nat. Photonics {\bf10}, 771 (2016).

\bibitem{19}	K. Krupa, A. Tonello, B. M. Shalaby, M. Fabert, A. Barthélémy, G. Millot, S. Wabnitz, and V. Couderc, Nat. Photonics {\bf11}, 237 (2017).

\bibitem{20}	O. S. Sidelnikov, E. V. Podivilov, M. P. Fedoruk, and S. Wabnitz, Opt. Fiber Technol. {\bf53}, 101994 (2019).


\bibitem{21}	E. V. Podivilov, D. S. Kharenko, V. A. Gonta, K. Krupa, O. S. Sidelnikov, S. Turitsyn, M. P. Fedoruk, S. A. Babin, and S. Wabnitz, Phys. Rev. Lett. {\bf122}, 3 (2019).

\bibitem{22}	E. Deliancourt, M. Fabert, A. Tonello, K. Krupa, A. Desfarges-Berthelemot, V. Kermene, G. Millot, A. Barthélémy, S. Wabnitz, and V. Couderc, OSA Contin. {\bf2}, 1089 (2019).

\bibitem{23}	J. Lægsgaard, Opt. Lett. {\bf43}, 2700 (2018).

\bibitem{24}	S. K. Dacha and T. E. Murphy, Optica {\bf7}, 1796 (2020).

\bibitem{25}	P. Mondal, V. Mishra, and S. K. Varshney, Opt. Fiber Technol. {\bf54}, 102041 (2020).

\bibitem{26}	K. Krupa, A. Tonello, A. Barthélémy, T. Mansuryan, V. Couderc, G. Millot, P. Grelu, D. Modotto, S. A. Babin, and S. Wabnitz, APL Photonics {\bf4}, (2019).

\bibitem{27}	O. Tzang, A. M. Caravaca-Aguirre, K. Wagner, and R. Piestun, Nat. Photonics {\bf12}, 368 (2018).

\bibitem{28}	F. O. Wu, A. U. Hassan, and D. N. Christodoulides, Nat. Photonics {\bf13}, 776 (2019).

\bibitem{29}	A. Fusaro, J. Garnier, K. Krupa, G. Millot, and A. Picozzi, Phys. Rev. Lett. {\bf122}, 123902 (2019).

\bibitem{30}	L. G. Wright, Z. M. Ziegler, P. M. Lushnikov, Z. Zhu, M. A. Eftekhar, D. N. Christodoulides, and F. W. Wise, IEEE J. Sel. Top. Quantum Electron. {\bf24}, 1 (2018).

\bibitem{31}	F. Poletti and P. Horak, J. Opt. Soc. Am. B {\bf25}, 1645 (2008).

\bibitem{32}	J. Milnor, Commun. Math. Phys. {\bf99}, 177-195 (1985).





\end{thebibliography}
\end{document}